\documentstyle[12pt]{article}
\def\vtd{$V_{td}$}

\def\lsim{\mathrel{\lower4pt\hbox{$\sim$}}
\hskip-12pt\raise1.6pt\hbox{$<$}\;}

\def\gsim{\mathrel{\lower4pt\hbox{$\sim$}}
\hskip-12pt\raise1.6pt\hbox{$>$}\;}

\begin{document}
\setlength{\baselineskip}{18pt}
\rightline{SLAC-PUB-6635}
\rightline{BNL-60709}
\rightline{TECHNION-PH-94-11}

\begin{center}

{\bf Feasibility of Extracting \vtd\ from Radiative $B(B_S)$ Decays}
\vspace{1cm}

{D. Atwood,$^{a)}$ B. Blok,$^{b)}$ and A. Soni$^{c)}$}
\vspace{1cm}

$a)$ Stanford Linear Accelerator Center, Stanford, CA\ \ 94309 \\
$b)$ Institute for Theoretical Physics, Technion, Haifa, Israel \\
$c)$ Brookhaven National Laboratory, Upton, NY\ \ 11973 \vspace{2cm}

\end{center}

\begin{quote}
{\bf Abstract} \qquad We use existing measurements of $D^-\to
K^{\ast0}\rho^-$ and $B\to \psi+K^\ast$, coupled with flavor
independence of QCD, and
with vector meson dominance to show that long distance
contributions to $B\to \rho +\gamma$ are potentially very serious. We
note that long distance (LD) contributions can be appreciably different
in $B^-\to \rho^-+\gamma$ and $B^0\to \rho^0 (\omega) +\gamma$. All
radiative decays of $B$, $B_S$ are shown to be governed essentially by
two LD and two short-distance (SD) hadronic entities. Separate
measurements of $B^-\to \rho^-+\gamma$, $B^0\to \rho^0 (\omega)
+\gamma$, along with $B\to K^\ast +\gamma$ appear necessary for a
meaningful extraction of \vtd. Measurements of $B_S\to \phi+\gamma$ and
$K^{\ast0}+\gamma$ could also provide very useful consistency checks.
\end{quote}

\section{Introduction}

The Cabibbo-Kobayashi-Maskawa (CKM) mixing angle \vtd\ is a parameter of
crucial importance to the Standard Model (SM) and it is still very
poorly known \cite{ckm}. Considerable experimental  effort is directed
towards its determination via the rare decay $K^+\to \pi^+\nu\bar\nu$
\cite{hl}. This process is considered to be theoretically clean for
extraction of \vtd\ \cite{buras}. However its branching ratio is
extremely rare being about a few times $10^{-10}$ rendering a precise
determination of \vtd\ very challenging. In $B$-physics one well known
method for determining \vtd\ is via the experimentally measured
$B$-$\bar B$ mixing. This requires a knowledge of the pseudoscalar decay
constant $f_B$ and the ``bag parameter'' $B_B$. Neither of these
quantities is directly accessible to experiment, at least not in the
near future. $f_B$ could eventually be measured directly in $B$ decays,
say via $B\to\tau+\nu_\tau$; but this will surely take a long time.
The reliability of the theoretical calculations for $f_B$ and $B_B$ may
therefore be a cause for concern. In any case the importance of \vtd\
demands that we determine it in many ways and with as much precision as
possible.

One $B$-decay in which \vtd\ enters is $B\to\rho+\gamma$
\cite{ali,soares,soni}. Since the related decay $B\to K^\ast+\gamma$ has
already been detected \cite{ammar} it is useful to understand what we
may learn about \vtd\ through a measurement of $B\to \rho+\gamma$. Rough
estimates indicate that LD contribution to $B\to\rho+\gamma$ are
potentially very serious. Since it is very difficult to accurately
estimate these LD contributions a precise extraction of \vtd\ from $B\to
\rho+\gamma$ \cite{bmeson} therefore also appears rather difficult.

In this paper we try to quantify various LD and SD sources for radiative
decays of all of the $B(B_S)$ meson, i.e.\ for:

\begin{eqnarray}
&B^-  \to\rho^-+\gamma & \qquad\qquad\qquad\qquad\qquad\qquad(1) \nonumber \\
&B^0  \to\rho^0+\gamma & \qquad\qquad\qquad\qquad\qquad\qquad(2) \nonumber \\
&B^0  \to\omega+\gamma & \qquad\qquad\qquad\qquad\qquad\qquad(3) \nonumber \\
&B^-  \to K^{\ast-}+\gamma & \qquad\qquad\qquad\qquad\qquad\qquad(4)
\label{btos}\\
&B^0  \to K^{\ast0}+\gamma & \qquad\qquad\qquad\qquad\qquad\qquad(5) \nonumber
\\
&B_S  \to\phi+\gamma & \qquad\qquad\qquad\qquad\qquad\qquad(6) \nonumber \\
&B_S  \to K^{\ast0} +\gamma & \qquad\qquad\qquad\qquad\qquad\qquad(7) \nonumber
\end{eqnarray}

\noindent We show that two types of LD and essentially two types of SD
contributions determine all of these decays. Thus separate experimental
measurements of as many of these reactions as possible could allow a model
independent determination of the hadronic entities and provide useful
self consistency checks. Consequently, extraction of \vtd\ to a
meaningful level of accuracy in the long run may become possible.
Clearly the necessary effort is then many times more than what is needed
for a single measurement of $B\to \rho+\gamma$. On the other hand, we
anticipate intense experimental activity in the area. Improvements at
existing $e^+e^-$ facilities such as CESR and LEP as well as
construction of new $e^+e^-$ based $B$-factories at SLAC and KEK will
lead to an increased sample of $B$'s. Furthermore many dedicated $B$
experiments are being proposed or planned at hadron machines. Bearing
all that in mind we give a general strategy for attempting to extract
\vtd\ precisely from radiative $B$-decays.

\section{A Close Look at $B\to\rho+\gamma$.}

\subsection{The Long Distance Contribution from $u\bar u$ States.}

It has been known for a long time \cite{hou} that for $b\to d$
flavor-changing loop transitions (unlike for $b\to s$) the tree graphs
(i.e.\ long-distance) become appreciably large and can easily dominate
over the loop (i.e.\ the SD) process. A simple example is the process

\begin{equation}
B^-\to d\bar u\gamma \to \rho^-\gamma \label{bminus}
\end{equation}

\noindent via the non-spectator (or the annihilation) mechanism shown in
Fig.~1a. Notice that this graph goes via $V_{ub}$, i.e.\ another poorly
known CKM parameter. So the reaction $B\to \rho+\gamma$ can occur, in
principle, even if \vtd\ is vanishingly small. Although it is very
difficult to accurately calculate such a contribution there are several
ways of estimating its size, i.e.\ within a factor of two or three. We
outline below two ways of calculating such contributions.

In the first method we invoke the correspondence of such annihilation
graphs with spectator plus final state interactions (FSI) to note that
Fig.~1(a) is exactly the same as Fig.~1(b). Fig.~1(b) shows the color
allowed simple spectator contribution to $B^-\to \rho^-+\rho^0_V$
followed by $\rho^0_V\to\gamma$ (where the subscript $V$ stands for
virtual). The first step of $B^-\to \rho^-+\rho^0$ can be estimated by
normalizing with the observed analogous decay: $D^-\to \rho^-+K^{\ast0}$
via Fig.~1(c). This correspondence between the two decays should hold
because of the flavor ($b\leftrightarrow c$) symmetry of QCD\null. To
the extent that $m_c(m_b) \gg \Lambda_{QCD}$ the effects of QCD do note
care about the flavor-label charm or bottom \cite{quarksym}.

Also SU(3) flavor symmetry ensures that the change from $K^{\ast0}$ to
$\rho^0$ in $D$ versus $B$ decay is mild apart from phase-space
correction. The conversion from $\rho\to\gamma$ can be dealt with by
using vector-meson dominance. Thus

\begin{equation}
\frac{BR(B^-\to \rho^-\rho^0)}{BR(D^-\to \rho^- K^{\ast0})} =
\left(\frac{m_b}{m_c}\right)^5 \left|\frac{V_{ub}}{V_{cs}} \right|^2
\frac{\tau_{B^+}}{\tau_{D^+}} \chi_{ps} \label{brbr}
\end{equation}

\noindent where $\chi_{ps}$ is the phase space ratio.

The conversion from $\rho^0$ to photon can be crudely estimated by using
VMD to amount a multiplicative factor of about $3\ast10^{-3}$
\cite{vmd,egsa}. Using $\left| \frac{V_{ub}}{V_{cb}} \right| = .08, |V_{cb}|
 = .04$ we thus find:

\begin{equation}
BR( B^-\to\rho^-\gamma)_{L_{u1}} = 6\times10^{-8} \label{br}
\end{equation}

\noindent where the subscript $L_u$ is to denote the LD contributions
coming from $u\bar u$ state(s) such as $\rho^0$.

A second method for estimating the same contribution is to use bound
state method of Ref.~13 for writing down the amplitude for $B^-\to\bar u
d\gamma$:

\begin{equation}
\begin{array}{c}
A(B^-\to \bar ud\gamma)_{1a} \simeq \frac{f_B}{16} e_ue
\frac{g^2_W}{m^2_W}\;\frac{1}{m_u} V_{ub} Tr \left[(\not p_B-m_B)
\not\epsilon_\gamma \not q \gamma^u P_L \right] \\ \\
\bar u_d \gamma^u P_Lv_u
\end{array} \label{ab}
\end{equation}

\noindent where $m_u$  is the constituent mass of the $u$ quark amd
$e_u$ is its charge (i.e.\ 2/3). Now  the light-quark current can be
(vacuum) saturated by the $\rho^-$ via use of:

\begin{equation}
\langle 0 |\bar u_d\gamma_\mu\gamma_5 v_u|\rho^-(p_\rho,\epsilon_p)\rangle =
f_\rho
m_\rho \epsilon^u_\rho \label{lang}
\end{equation}

\noindent where $f_\rho$ is the decay constant of $\rho$ i.e.\ about
220 MeV\null. Thus:

\begin{equation}
\frac{\Gamma(B^-\to\rho^-\gamma)_{1a}}{\Gamma(B^- \to\bar
ud\gamma)_{1a}}  \simeq  18\pi^2 \frac{f^2_\rho m^2_\rho}{m^4_B}
\left( 1-\frac{m^2_\rho}{m^2_B} \right) \simeq 7\times10^{-3}  \label{Gamma}
\end{equation}

\noindent Using (\ref{ab}) and (\ref{Gamma}) we arrive at a second estimate
for the LD correction due to $\bar uu$ states

\begin{equation}
BR (B^-\to \rho^-\gamma)_{L_{u2}}  \simeq  8\times 10^{-8} \label{brb}
\end{equation}

\noindent where we have used $f_B =180$ MeV \cite{bernard} and $m_u=330$
MeV\null.
In passing we note from eqns.~\ref{ab}--\ref{Gamma} that the inclusive
branching ratio for the reaction $B^-\to \bar u d\gamma$ via the
annihilation graph is given by:

\begin{equation}
BR(B^-\to u\bar d \gamma) \approx 1.1\times 10^{-5} \label{brbm}
\end{equation}

Given the intrinsic uncertainties in each of the two methods outlined
above the resulting numbers in eqns.~(\ref{br}) and (\ref{brb}) should be
regarded as
in qualitative agreement. Thus for one class of long distance
contributions, namely those due to $u\bar u$ states we will take the
mean of the two numbers from eqn.~(\ref{br}) and eqn.~(\ref{brb}) and rather
arbitrarily assign a factor of four uncertainty. Thus for the
corresponding amplitude we get

\begin{equation}
A(B^-\to \rho^-\gamma)_{L_u} = (2-4) \times 10^{-4} \label{abminus}
\end{equation}

Now let us address the case of the neutral $B$ i.e.\ the corresponding
LD contributions from $\bar uu$ states to ($B^0\to\rho^0\gamma$). Then
Fig.~1(a) gets redrawn as Fig.~1(d) and Fig.~1(b) gets redrawn as
Fig.~1(e). In each case we see that the graphs for $B^0$ are color
suppressed. Thus

\begin{eqnarray}
A(B^0\to\rho^0\gamma)_{L_u} & = & -\left(\frac{1}{3} \times\frac{3}{2} \times
\frac{1}{\sqrt{2}} \right) \times (2-4) \times 10^{-4} \label{abzero}  \\
A(B^0\to\omega\gamma)_{L_u} & = & \frac{1}{6\sqrt{2}} (2-4) \times 10^{-4}
\label{abzeronew}
\end{eqnarray}

\subsection{The Long Distance Contributions from $c\bar c$ States.}

We next turn our attention to the LD contributions to
$B^-\to\rho^-+\gamma$ from $c\bar c$ states. The most notable origin is
the chain $B^-\to\rho^-+\psi_V$ followed by $\psi_V\to\gamma$. Using the
measured rate

\begin{equation}
BR(B^0\to K^{\ast0}\psi) = (1.6\pm.3) \times 10^{-3} \label{brbz}
\end{equation}

\noindent we immediately get

\begin{equation}
BR(B^-\to \rho^-\psi) = 2 BR(B^0\to\rho^0\psi) = \lambda^2  Br(B^0\to
K^{*0}\psi) p_{s K^\ast\rho} \label{brbp} \end{equation}

\noindent where $\lambda\equiv \sin\theta_c=.22$ and $p_{s K^\ast\rho}$
is a phase space correction factor estimated to be about 1.4 due to the
mass difference between $\rho$ and $K^\ast$ \cite{formulas}. Following
Ref.~11 conversion factor from $\psi\to\gamma$ is estimated at
$5\times10^{-3}$. However, in this $\psi\to\gamma$ conversion we want
to include only the transversely polarized fraction of $\psi$'s. These
are estimated to be about 30\% \cite{alamcleo}. Thus, for the amplitude
of LD contributions  from $c\bar c$ states we get

\begin{equation}
A(B^-\to\rho^-\gamma)_{L_c} = (2-6)\times 10^{-4} \label{abplus}
\end{equation}

\noindent where in specifying the range we are again estimating about a
factor of two uncertainty (in the amplitude).

\subsection{The Short-Distance Contributions to $B\to\rho+\gamma$}

The SD (or penguin) contributions arise from loop graphs, such as
Fig.~1(f) and 1(g). It is known for a long time that QCD corrections
play an important role here. We recall that this is due to the fact that
in the pure electroweak penguin (Fig.~1(f)) there is an accidental
cancellation of the coefficients of terms that maintain GIM unitarity
with a logarithmic dependence on the internal quark mass (i.e.\ $m_u$,
$m_c$, $m_t$). As a result the leading terms exhibit a power law
dependence on that mass. On switching on QCD the coefficient of the log
term becomes nonvanishing and results in enhanced QCD radiative effects.

By now there is an extensive literature describing the effects of QCD on
radiative decays of $B$'s. For our purpose it is useful to first discuss
the $b\to s$ process namely the one relevant to $B^-(B^0)\to
K^{\ast-}(K^{\ast 0}) +\gamma$ (or to $B_S\to\phi+\gamma$). Recall the CKM
unitarity for this channel:

\begin{equation}
v^q_u + v^q_c + v^q_t =0 \label{lcv}
\end{equation}

\noindent where $v^q_j=V_{jb}V^\ast_{jq}$, $j=u,c,t$ and $q=s$ or $d$.
Recall also that \cite{ckm}

\begin{eqnarray}
V_{us}  = & \lambda & \simeq\; .22 \label{vus} \\
\frac{V_{ub}}{V_{cb}}  = & .08 & \pm\; .03 \label{vubcb} \\
V_{tb}  = &  .99 & \pm\; .01 \label{vtb}
\end{eqnarray}

\noindent Thus for $b\to s$ case the up quark term ($V_{ub}V^\ast_{us}$)
is negligible in comparison to the other two terms in
eqn.~(\ref{lcv}).
This has two important consequences. First is that one gets the usual
relation:

\begin{equation}
V_{ts} \simeq -V_{cb} \label{vts}
\end{equation}

\noindent to a very good approximation. The second important consequence
of the smallness of $V_{ub}V^\ast_{us}$ is that in the $b\to s$
penguin loop the $u$ quark contribution is forced to become so small
that the precise dependence on $m_u$ is not at all important. Such is
not the case for $b\to d$ penguins as we will soon elaborate.

The penguin (SD) contributions can be written as

\begin{equation}
A^q_p = \sum_j f_j v^q_j \label{aqp}
\end{equation}

\noindent For $q=s$, we can use
eqn.~(\ref{lcv})
and rewrite

\begin{equation}
A^s_p = (f_t-f_c) v^s_t + (f_u-f_c) v^s_u \label{asp}
\end{equation}

\noindent Since $v^s_u$ is extremely small the second term is bound to
make a negligibly small contribution and consequently the assumption
that $f_c=f_u$ that one usually makes \cite{burastwo} becomes a very safe
assumption. Then for $b\to s$ with a very good approximation one gets

\begin{equation}
A^s_p = (f_t-f_c) v^s_t \equiv (f_t-f_c) V_{ts} \label{asptwo}
\end{equation}

\noindent For the case of $b\to d$ transitions the $u$ quark in the loop
no longer appears with the small parameter $V_{us}(\equiv\lambda)$
multiplying its effects and the charm and the top quark both now have
smaller CKM factors monitoring their contributions to the penguin
amplitude. The $u$ quark contribution is no longer necessarily
negligible in comparison to the others and the assumption $f_c=f_u$ is
no longer a good approximation since it forces a potentially important
($u$ quark) contribution to unnaturally vanish. Any reasonable deviation
of $f_c/f_u$ away from unity would have important corrections. To make
the best use of the experimental information that one gets from
measurement of $B\to K^\ast\gamma$, it is prudent now to use unitarity
and rewrite the $b\to d$ penguin as:

\begin{equation}
A^d_p = (f_t-f_c) v^d_t + (f_u -f_c) v^d_u \label{adp}
\end{equation}

\noindent Taking ratios of equations~(\ref{asptwo}) and (\ref{adp}):

\begin{equation}
\begin{array}{rl}
A^d_p/A^s_p  & = \frac{V_{td}}{V_{ts}} [ 1+ \Delta] \label{adpasp_a}\\
\ & \ \\
\Delta & \equiv \left(\frac{f_u-f_c}{f_t-f_c} \right)
\left(\frac{V_{ub}}{V_{td}} \right)
\end{array} \label{adpasp_b}
\end{equation}

\noindent Thus there are two hadronic entities:

\begin{eqnarray}
f_u-f_c & \equiv & S_{uc} \label{fufc} \\
f_t-f_c & \equiv & S_{tc} \label{ftfc}
\end{eqnarray}

\noindent monitoring all the SD contributions in $b\to s$ and $b\to d$
penguins. $f_t$ and $f_c$ have recently been calculated in Ref.
\cite{burastwo}:

\begin{eqnarray}
f_t & \simeq & -.11 \label{ft} \\
f_c & \simeq & .16 \label{fc}
\end{eqnarray}

\noindent giving

\begin{equation}
S_{tc} =-.27 \label{stc}
\end{equation}

For extraction of \vtd\ from experiment the deviation from unity of the
quantity in square parenthesis in equation (\ref{adpasp_a})
is important. First let us estimate the CKM
ratio that enters there. We note that the use of \cite{bernard}

\begin{eqnarray}
f_B & = & 180 \pm 40 \mbox{ MeV} \label{fbeq} \\
B_B & = & 1\pm.2 \label{bbeq}
\end{eqnarray}

\noindent emerging from lattice calculations along with the measured
$B$-$\bar B$ mixing gives

\begin{equation}
\frac{V_{td}}{V_{ts}} \simeq 0.22\pm.08 \label{vtdts}
\end{equation}

\noindent Thus using (as 90\% CL bounds)

\begin{eqnarray}
|V_{ub}/V_{cb}| & < & .13 \label{vubvcb} \\
|V_{td}/V_{ts}| & \ge & .09 \label{vtdvts}
\end{eqnarray}

\noindent we get

\begin{equation}
|V_{ub}/V_{td}| \lsim 1.5 \label{vubvtb}
\end{equation}

A precise value for $f_u-f_c$ is extremely difficult to calculate. We
will assume that $f_u$ and $f_c$ depend logarithmically on $m_u$ and
$m_c$. Using constituent masses $m_u\simeq0.3$ GeV, $m_c\simeq1.8$ GeV
and the numerical result (\ref{fc}) of Ref.~\cite{burastwo} we then {\it
crudely
estimate}:

\begin{equation}
S_{uc}/S_{tc} = -.30 \label{suc}
\end{equation}

\noindent Thus the ratio of the SD amplitudes for $b\to d$ and $b\to s$
may deviate appreciably from the CKM ratio $V_{td}/V_{ts}$. We note this
deviation from the simple CKM scaling is controlled crucially by the
ratio $V_{ub}/V_{td}$ just as the relative importance of the LD
contributions due to $u\bar u$ states (i.e\ $L_u$) to $B\to\rho+\gamma$
is controlled by $V_{ub}/V_{td}$. If the mild indications from the
current central values of $V_{ub}/V_{cb}$ and $V_{td}/V_{ts}$ (.08
versus .22)
is confirmed then the extraction of \vtd\ from
$B\to\rho+\gamma$ will indeed be easier than otherwise.

To gauge the relative importance of the LD and the SD contributions to
($B\to\rho+\gamma$) we need to estimate $A^s_p$ (i.e.\ SD amplitude for
$b\to s$) so as to be able to use eqn.~(\ref{adpasp_a}) to get $A^d_p$
(i.e.\ the SD amplitude for $b\to d$). We can try to use the
experimental result on $B\to K^\ast+\gamma$ for that purpose; so we turn
our attention to that reaction now.

\subsection{The Long- and Short Distance Contributions to $B\to
K^\ast+\gamma$}

The LD contribution from $\bar uu$ states is easily estimated from
eqn.~(\ref{abminus})

\begin{eqnarray}
A^s(B^-\to K^{\ast-}\gamma)_{L_u} &  \simeq & (4-8)\times10^{-5}
\label{asbm} \\
A^s(B^0\to K^{\ast0}\gamma)_{L_u} & \simeq & (2-4)\times 10^{-5} \label{asbz}
\end{eqnarray}

\noindent Similarly, from eqn.~(\ref{brbz}), with use of the
$\psi\to\gamma$ conversion factor of $5\times10^{-3}$ and incorporating
a factor of 0.3 for the fraction of transversely polarized $\psi$'s we
get

\begin{equation}
A(B\to K^\ast+\gamma)_{L_c} = (1-3)\times10^{-3} \label{abtokast}
\end{equation}

\noindent So for $B\to K^\ast\gamma$ the LD contributions due to $c\bar
c$ completely dominate over the $u\bar u$ ones \cite{another}.

Recall now the recent experimental result \cite{ammar}

\begin{equation}
BR(B\to K^\ast\gamma) = (4.5\pm1.5\pm 0.9) \times 10^{-5} \label{bbtokast}
\end{equation}

\noindent For the amplitude we translate this as

\begin{equation}
A(B\to K^\ast\gamma)\mid_{\rm expt} \simeq (6.7\pm1.7) \times 10^{-3}
\label{abtokastgam}
\end{equation}

\noindent From equations (\ref{abtokast}) and (\ref{abtokastgam}) we see
that there can be about 15--50\% LD contributions in the observed
experimental result. Combining those two equations we arrive at the SD
component

\begin{equation}
A^s_p \equiv A(B\to K^\ast+\gamma)_{SD} = (4.7\pm2.7) \times 10^{-3}
\label{aspeq}
\end{equation}

\noindent In arriving at eqn.~(\ref{aspeq}) we have made a strong
assumption that the SD and LD ($c\bar c$) amplitudes for $B\to
K^\ast+\gamma$ have the same relative sign. This assumption is based on
the belief that an opposite choice of signs would make the exclusive to
inclusive ratio for the short distance component alone, i.e.

\begin{equation}
H_{K^\ast} = \frac{\Gamma(B\to K^\ast+\gamma)}{\Gamma(b\to s+\gamma)}
\label{gamovgam}
\end{equation}

\noindent become uncomfortably large. The point is that lattice methods
have been used to calculate this hadronization ratio for the single
operator

\begin{equation}
\bar s_L \sigma_{\mu\nu}b_R F^{\mu\nu} \label{barsl}
\end{equation}

\noindent that emerges from the short distance expansion. The results of
the lattice calculation are \cite{cbernard}:

\begin{equation}
H_{K^\ast} = 6.0\pm 1.2\pm 3.4\%  \label{hkast}
\end{equation}

\noindent For our purpose we will adopt a very conservative
interpertation of the lattice results, namely

\begin{equation}
H_{K^\ast} < 12\%  \label{hkasttwo}
\end{equation}

Recall now the recent CLEO result \cite{cleoincl}:

\begin{equation}
BR(b\to\gamma+s) = (2.32\pm.51\pm.29\pm.32) \times 10^{-4}  \label{cleo}
\end{equation}

\noindent Combining equations (\ref{bbtokast}) and (\ref{cleo})
indicates that the experimental value of the exlusive to inclusive ratio
is around 20\% which tends to be on the high side compared to the
lattice expectation. By attributing a fraction of the observed exclusive
signal to come from LD sources as in equation~(\ref{aspeq}) brings the
hadronization ratio for the SD piece i.e.

\begin{equation}
\frac{(4.7\times10^{-3})^2}{2.3\times10^{-4}} \sim 9.6\%
\label{sdpiece} \end{equation}

\noindent to be more in the ball park of the lattice results. If, on the
other hand, we take the LD and SD contributions to $B\to K^\ast\gamma$
to have a relative negative sign then the SD fraction would have to be

\[ \frac{(8.7\times10^{-3})^2}{2.3\times10^{-4}} \sim 33\% \]

\noindent which is too large from the lattice persepective.

\subsection{Estimates for the Relative Importance of LD
Contribution to $B\to \rho+\gamma$}

Using eqn.~(\ref{adpasp_a}) and (\ref{aspeq}) and invoking SU(3) gives us
the SD contribution to ($B\to \rho+\gamma$)

\begin{eqnarray}
A(B\to\rho+\gamma)_{SD} & = & \frac{V_{td}}{V_{ts}} [1+\Delta] \times
(4.7\pm2.7)\times10^{-3} \label{abtorho} \\
& \sim & (5-15)\times10^{-4} \label{sim}
\end{eqnarray}

\noindent From eqn.~(\ref{abplus}) and (\ref{sim}) we see that for
$B^-\to \rho^-+\gamma$ the  LD $c\bar c$ states are at least 15\% of
(and could even dominate over) the SD ones. Indeed even that minimum
value of 15\% implies a contamination of these LD effects on the rate
for $B^-\to\rho^-\gamma$ to approach 30\%. From eqn.~(\ref{abminus}) we
see that the $u\bar u$ states seem to be somewhat less important than
the $c\bar c$ but are roughly comparable. We emphasize again that the
numbers given for $L_u$ in eqn.~\ref{abminus} assume
$\left|\frac{V_{ub}}{V_{cb}} \right| = .08$. Given the intrinsic
difficulties of the LD estimates it appears doubtful that
$B^-\to\rho^-+\gamma$ alone in conjunction with $B\to K^\ast+\gamma$ can be
used to deduce reliable information on \vtd\ before a lot more experimental
data on radiative decays becomes available. In this regard a precise
value of $V_{ub}$ as well as the relative sign of $V_{ub}V^\ast_{ud}$
with $V_{cb}V^\ast_{cd}$ is very important since a relative negative
sign between these two CKM elements would result in (at least) a partial
cancellation of the long distance $L_u$ and $L_c$ terms.

The LD contribution to $B^0\to\rho^0+\gamma$ from $u\bar u$ states are
substantially less (see eqn.~(\ref{abzero})) than for
$B^-\to\rho^-+\gamma$. The SD contributions are the same for $B^0$ and
$B^-$ (i.e.\ eqns.~(\ref{abtorho}) and (\ref{sim})). Thus
$B^0\to\rho^0+\gamma$ may have appreciable advantages over
$B^-\to\rho^-+\gamma$ for learning about \vtd. In any event, it seems
clear from the preceding estimates that the rates for
$B^-\to\rho+\gamma$ may be quite different from that of
$B^0\to\rho+\gamma$. Since the SD contributions (which scale with
\vtd) are the same for $B^-$ and $B^0$ and the LD ones are not, separate
measurements of $B^-$ and $B^0$ radiative decays are important to
understanding the dynamics of these decays and
they are essential
for facilitating any
reliable determination of \vtd.

\section{Four Hadronic Entities Essentially Determine all the Radiative
$B$-Decays.}

In the preceding section we have discussed the long and short distance
contributions to charged and neutral $B$ decays to $\rho+\gamma$ and
$K^\ast+\gamma$. During the course of that discussion we had to
introduce two LD (namely $L_u$ and $L_c$) and two short distance (namely
$S_{tc}$ and $S_{uc}$) entities. Indeed all the radiative $B$, $B_S$
decays to the seven final states given in eqn.~(\ref{btos}) are governed
by the same four hadronic entities \cite{blok}. Of course the dependence on CKM
angles are not the same (also there are obvious differences in $N_c$
dependence and on flavor SU(3))  that have to be taken into account.
Thus we can write

\begin{equation}
A(B^-\to\rho^-+\gamma)  =  e_u \left[
\left( N_C - 1 \right)
v^d_u
L_u + v^d_c{L_c} + k_bc_{B\rho}T_{1_{B\rho}} (v^d_t S_{tc} + v^d_u S_{uc})
\right]
\label{abptorho}
\end{equation}
\begin{equation}
A(B^0\to\rho^0+\gamma)\! =\!  -\frac{1}{\sqrt{2}} \left[ (e_u-e_d) v^d_u
{L_u} + e_u  v^d_c {L_c} + k_bc_{B\rho}T_{1_{B\rho}} \left( v^d_t S_{tc}
+ v^d_u S_{uc} \right) \right]  \label{abztorho}
\end{equation}

\noindent Also

\begin{equation}
A(B^0\to \omega+\gamma) \! =\! \frac{1}{\sqrt{2}} \left[ (e_u+e_d) v^d_u
{L_u} + e_u v^d_c {L_c} + k_bc_{B\rho}T_{1_{B\rho}} \left( v^d_t S_{tc}
+ v^d_u S_{uc} \right)\right] \label{abztow}
\end{equation}
\begin{eqnarray} A(B^-\to K^{\ast-}+\gamma) & = & e_u \left[ v^s_u
N_c L_u + v^s_c {L_c} + k_bc_{BK^\ast}T_{1_{BK^\ast}} (v^s_t S_{tc} +
v^s_u S_{uc}) \right] +\nonumber \\
& \simeq & e_u \left[ v^s_c {L_c} + k_bc_{BK^\ast}T_{1_{BK^\ast}}
v^s_t S_{tc} \right] \label{abptok} \\
A(B^0\to K^{\ast0}+\gamma) & = & e_u \left[ v^s_u L_u +  v^s_c {L_c} +
k_bc_{BK^\ast}T_{1_{BK^\ast}}  (v^s_t S_{tc} + v^s_u S_{uc}) \right]
\nonumber \\
& \simeq & e_u \left[ v^s_c {L_c} +
k_bc_{BK^\ast}T_{1_{BK^\ast}}  v^s_t S_{tc} \right] \label{abztok}
\end{eqnarray}

\noindent Similarly for related decays of $B_S$:

\begin{eqnarray}
A(B_S\to\phi+\gamma) & = & e_u \left[ v^s_u L_u +
v^s_c L_c + k_bc_{B_s\phi}T_{1_{B_s\phi}} (v^s_t S_{tc} + v^s_u S_{uc})
\right] \nonumber \\
& \simeq & e_u \left[ v^s_c L_c + k_bc_{B_s\phi}T_{1_{B_s\phi}} v^s_t S_{tc}
\right] \label{abstophi} \\
A(B_S\to K^\ast+\gamma)\!\! & = &\!\! e_u \left[ v^d_u L_u + v^d_c L_c +
k_bc_{B_sK^\ast}T_{1_{B_sK^\ast}} (v^d_t
S_{tc} + v^d_u S_{uc}) \right] \label{abstok}
\end{eqnarray}

\noindent Here $k_b$ is a normalization constant designed so that the
width for the flavor-changing transition coming from the short distance
piece alone is related properly to the factors $S_{tc}$ and $S_{uc}$.
Thus

\begin{equation}
\Gamma(b\to d\gamma)_{SD} \equiv\Gamma(b\to d\gamma)_{\rm penguin} =
[ e_uk_b (v^d_tS_{tc} + v^d_uS_{uc})]^2 \label{penguin}
\end{equation}

\noindent $T_1$ is the only form factor (at $q^2=0$) that determines the
exclusive to inclusive ratio from the short-distance penguin part
\cite{another}. Thus

\begin{equation}
\frac{\Gamma(B\to \gamma\rho)_{SD}}{\Gamma(b\to\gamma d)_{SD}} =
c^2_{B\rho}T^2_{1B\rho} \label{pengpart}
\end{equation}

\noindent where

\begin{equation}
C^2_{B\rho} = 4 \left( \frac{m_B}{m_b}\right)^3 \left[ 1-
\frac{m^2_{\rho}}{m^2_B} \right]^3 \label{cbrho}
\end{equation}

\section{Discussion}

In Table~1 we give rough estimates for the radiative modes. For
simplicity we have assumed that $T_1(q^2=0)$ is the same for $B\to
K^\ast,\rho$ and $B_s\to K^\ast, \phi$. There could easily be
differences between these form factors amounting to 10 or even 20\%.
Future lattice and QCD sum rule calculations should be able to determine
these quite reliably.

\begin{table}
\begin{center}
\caption{Numerical Estimates}
\medskip
\begin{tabular}{|l|c|c|c|c|c|}
\hline
Reaction &\multicolumn{3}{|c|}{$|\mbox{Amplitudes}|/10^{-4}$} &
\multicolumn{2}{|c|}{Branching Ratio/$10^{-7}$} \\
\cline{2-6}
&$(u\bar u)_{LD}$ & $(c\bar c)_{LD}$ & SD & SD only & Total \\
\hline
$B^+\to\rho^+\gamma$ & $3\pm1$ & $4\pm2$ & $10\pm6$ & 2--25 & .4--68 \\
$B^0\to\rho^0\gamma$ & $1.1\pm.4$ & $2.8\pm1.4$ & $7\pm4$ & 1--12 & 1--32 \\
$B^0\to\omega^0\gamma$ & $.2\pm.1$ & $2.8\pm1.4$ & $7\pm4$ & 1--12 & 2--23 \\
$B\to K^\ast\gamma$ & $.6\pm.3$ & $20\pm10$ & $50\pm30$ & 40--640 & 90--1200 \\
$B_S\to K^\ast\gamma$ & $1.5\pm.5$ & $4\pm2$ & $10\pm6$ & 2--26 & 2--58 \\
$B_S\to\phi\gamma$ & $.6\pm.2$ & $20\pm10$ & $50\pm30$ & 40--640 & 90--1200 \\
\hline
\end{tabular}
\end{center}
\end{table}

Notice that the spread in the range due to the SD piece alone is less
than the spread after the LD contributions are included. This is in part
because the relative signs are not known at this time. Thus typically the
SD piece alone has a range of about one order of magnitude and that
increases appreciably to the extent that in one case it becomes as much
as two orders of magnitude when the LD pieces are also included.

We must also emphasize that the entries in the table are highly
correlated so that as better experimental information on any mode(s)
becomes available then it will effect the estimates for all of the
modes. This is of course another way of saying that all of the decays
involve only a few (i.e.\ four) hadronic quantities. In the case of
$B\to K^\ast \gamma$ the recently obtained experimental measurement has
been used  to fix the relative sign between the SD piece and the LD
($c\bar c$) piece. There still remains an uncertainty in the theoretical
expectation for the total $BR$ of about one order of magnitude.
Measurements of $B\to
\rho(\omega)+\gamma$, especially separate ones for charged and neutral, will
significantly aid such an analysis in the future. Differences in
the $BR$s for $\rho^-+\gamma$, $\rho^0+\gamma$ and $\omega+\gamma$ would be
an excellent indicator of the extent of the LD contamination. If the LD
contributions are small then the $BR$s for these modes should follow the
expected factor of two difference due to the difference in their naive
quark content.

Indeed from eqns.~(\ref{abptorho}--\ref{abztow}) one finds:

\begin{eqnarray}
|A(B^-\to\rho^-+\gamma)| -\sqrt{2} |A(B^0\to\rho^0+\gamma)| & = &
V^d_u L_u [3e_u-e_d] \label{abmrhoone} \\
|A(B^-\to\rho^-+\gamma)| -\sqrt{2} |A(B^0\to \omega+\gamma)| & = & V^d_u L_u
[3e_u+e_d]. \label{abmrhotwo}
\end{eqnarray}

\noindent Thus experimental determination of the differences in the BR's
can be used to quantitatively deduce the long distance piece due to
$u\bar u$.

Lattice calculations of $B\to K^\ast+\gamma$ could also play a very
useful role. If improved lattice calculations for $B\to K^\ast\gamma$
also do not agree in their determination of the ratio $H_{K^\ast} \equiv
[ BR(B\to K^\ast+\gamma)/BR(b\to s+\gamma)]$ with improved experimental
measurements then the difference between the two must be attributed to
long distance pieces (presumably due to $c\bar c$ states) that the
lattice calculations do not include.
\bigskip

\noindent {\Large\bf Acknowledgements:}
\medskip

We are grateful to Ahmed Ali, Ikaros Bigi, Nilendra Deshpande, Bill
Marciano, Hubert Simma, Sheldon Stone and Ed Thorndike for discussions.
Work of D.A. and A.S. was supported in part by the US-Israel Binational
Science Foundation Grant and DOE contracts DE-AC03-76SF00515 and
DE-AC02-76CH0016. Work of B.B. was supported in part by the Israel
Science Foundation administered by the Israel Academy of Sciences and
Humanities and Technion V.P.R. fund. \bigskip

\noindent {\Large\bf Figure Captions}
\medskip

\begin{description}

\item[Fig.~1 a--e] A partial set of long distance contributions due to
$u\bar u$ states. Those due to $c\bar c$ states typically result by replacing
$u\to c$ in Fig.~1e.

\item[Fig.~1 f--g] Show typical penguin (short-distance) contributions.

\end{description}


\begin{thebibliography}{99}
\bibitem{bmeson} We will use the notation $B$ when we mean $B^\pm$
and/or $B^0(\bar B^0)$, i.e.\ when we do not mean to make a distinction
between charged and neutral $B$-mesons.
\end{thebibliography}
\end{document}